\documentclass{article}

\usepackage[utf8]{inputenc} 
\usepackage{hyperref}       
\usepackage{url}            
\usepackage{booktabs}      
\usepackage{amsfonts}       
\usepackage{nicefrac}       
\usepackage{microtype}     
\usepackage[english]{babel}
\usepackage{lipsum}
\ifpdf\usepackage{epstopdf}\fi
\usepackage{titlesec}
\titlelabel{\thetitle.\quad}
\usepackage{amsmath}
\usepackage{indentfirst}
\usepackage{amssymb}
\usepackage{physics}
\usepackage[a4paper, mag=1000, includefoot, left=3cm, right=2cm, top=2cm, bottom=2cm, headsep=1cm, footskip=1cm]{geometry}
\usepackage{caption}
\title {\textbf{\textit{CP}-violation in oscillations of three neutrino generations: the case of degenerate masses}}
\numberwithin{equation}{section}

\author{
 \textbf{E.K. Karkaryan\footnote{karkaryan@bk.ru}} \\
  \textit{NRC "Kurchatov Institute" - ITEP, Moscow, Russia} \\
}
\date{}
\begin{document}
\maketitle

\begin{abstract}
\textit{CP}-symmetry violates in oscillations of neutrinos in vacuum in presence of three generations. In this case there is an inherent complex phase in PMNS-matrix. When any two neutrino masses degenerate the \textit{CP}-violation vanishes. In the present article different mechanisms of \textit{CP}-symmetry recovery are demonstrated in case of two degenerate neutrino mass-states. 
\end{abstract}

\section{Introduction}
\setlength{\parindent}{5ex}
For neutrino oscillations the necessary condition of \textit{CP}-violation is the presence of the phase in the mixing matrix, which cannot be removed with the rotation of lepton fields in the Standard model. Only with the presence of three neutrino generations, the corresponding phase remains in PMNS-matrix. However when any two neutrino states are degenerate the \textit{CP}-symmetry recovers. This follows from the expression for the Jarlskog determinant lepton analog, which is proportional to the neutrino mass difference. Mostly in the literature the question of how the \textit{CP}-violation vanishes in case of degenerate masses is not considered in sufficient detail or is not raised at all. The present work is aimed to help the people who are engaged in the neutrino physics to understand this issue. There are two different explanations of \textit{CP}-symmetry recovery are considered in the article. The first demonstrates the explicit opportunity of removing the \textit{CP}-violating phase from the mixing matrix. Such an opportunity is covered in the literature on neutrino oscillations. The second mechanism, which differs in principle from the first one relies on the fact that the amplitude of oscillations depends on not only the \textit{CP}-noninvariant phase, but also \textit{CP}-invariant phase. In order for \textit{CP}-symmetry to be broken, the presence of different both \textit{CP}-preserving and \textit{CP}-violating phases in the amplitude of oscillations. So even at the presence of \textit{CP}-noninvariant phases in the amplitude the equality of \textit{CP}-invariant phases leads to the equality of probabilities of the straight and \textit{CP}-inverse processes, and consequently the \textit{CP}-symmetry recovery. This mechanism is not described in the literature. For the purposes of this article the detailed derivation of the amplitude and the neutrino oscillation probability in case of three neutrino is placed in the Appendix A.  

\section{Neutrino Oscillations in case of three generations}
\setlength{\parindent}{5ex}
The presence of the oscillations is the consequence of nonzero neutrino masses. Considering, that there is a neutrino with the momentum $\overrightarrow{\textbf{p}}$ in the initial state, and by considering the case of three neutrino generations $\nu_e, \nu_{\mu}, \nu_{\tau}$, we will obtain the neutrino oscillation probability at the distance $L$ from the source. Flavour states have no definite mass and represent the superposition of $\nu_1, \nu_2, \nu_3$ mass states. The neutrino mixing is described with the three-dimensional unitary matrix $U_{\mathrm{PMNS}}$:

\begin{equation}
\begin{pmatrix}
\nu_e\\
\nu_{\mu}\\
\nu_{\tau}
\end{pmatrix}=
\begin{pmatrix}
U_{e1} & U_{e2} & U_{e3} \\
U_{\mu 1} & U_{\mu 2} & U_{\mu 3} \\
U_{\tau 1} & U_{\tau 2} & U_{\tau 3}
\end{pmatrix}
\begin{pmatrix}
\nu_1\\
\nu_2\\
\nu_3
\end{pmatrix}.
\end{equation}

Parameterization of this matrix can be performed the same way as CKM-matrix \cite{cit1} by introducing three mixing angles $\theta_{12}, \theta_{23}, \theta_{13}$ and the phase $\delta$, which can not be eliminated by the transformation of the charged lepton and neutrino fields. Then $U_{\mathrm{PMNS}}$ will be explicitly of the form:

\begin{equation}
U_{\mathrm{PMNS}}= \label{eq2}
\begin{pmatrix} 
c_{13}c_{12} & s_{12}c_{13} & s_{13}e^{-i\delta}\\
-s_{12}c_{23}-c_{12}s_{13}s_{23}e^{i\delta} & c_{12}c_{23}-s_{12}s_{13}s_{23}e^{i\delta} & s_{23}c_{13}\\
s_{12}s_{23}-c_{12}c_{23}s_{13}e^{i\delta} & -c_{12}s_{23}-s_{12}c_{23}s_{13}e^{i\delta} & c_{23}c_{13} 
\end{pmatrix},
\end{equation}
where $c_{12}\equiv \cos{\theta_{12}}, s_{12}\equiv \sin{\theta_{12}}$ etc., $\delta$ is a Dirac \textit{CP}-violating phase.

The expression for neutrino oscillations with the flavour $\alpha$, energy $E$ and at the distance $L$ from the source in case of three neutrino generations is of the form:
\begin{equation}
\begin{gathered}
P_{\nu_{\alpha}\rightarrow{\nu_{\beta}}}=\delta_{\alpha\beta}-2\sum_{k>j}\Re{U_{\alpha k} U^{*}_{\beta k} U^{*}_{\alpha j} U_{\beta j}}(1-\cos{\frac{\Delta m^2_{kj}}{2E}}L)-2\sum_{k>j}\Im{U_{\alpha k} U^{*}_{\beta k} U^{*}_{\alpha j} U_{\beta j}}\sin{\frac{\Delta m^2_{kj}}{2E}L}=\\=\delta_{\alpha \beta}-4\sum_{k>j}\Re{U_{\alpha k} U^{*}_{\beta k} U^{*}_{\alpha j} U_{\beta j}}\sin^2{\frac{\Delta m^2_{kj}}{4E}L}-2\sum_{k>j}\Im{U_{\alpha k} U^{*}_{\beta k} U^{*}_{\alpha j} U_{\beta j}}\sin{\frac{\Delta m^2_{kj}}{2E}L},
\end{gathered}   
\end{equation} 
where  $U_{\alpha j}$ is the element of the mixing matrix, $\Delta m^2_{kj}\equiv m^2_k-m^2_j$. The derivation of this formula is given in the Appendix A.

Under \textit{CP}-transformation the process $\nu_{\alpha}\rightarrow{\nu_{\beta}}$ goes into the process $\overline{\nu_{\alpha}}\rightarrow\overline{\nu_{\beta}}$. The fact that the difference $\Delta P=P_{{\overline{\nu_{\alpha}}}\rightarrow{\overline{\nu_{\beta}}}}-P_{\nu_{\alpha}\rightarrow{\nu_{\beta}}}$ is nonzero is an evidence of the \textit{CP}-violating. The calculation of the value $P_{{\overline{\nu_{\alpha}}}\rightarrow{\overline{\nu_{\beta}}}}$ is similar to the calculation of $P_{\nu_{\alpha}\rightarrow{\nu_{\beta}}}$ with the only difference that for antiparticles complex-conjugated elements of mixing matrix are used:

\begin{equation}
P_{\overline{\nu_{\alpha}}\rightarrow{\overline{\nu_{\beta}}}}=\delta_{\alpha \beta}-4\sum_{k>j}\Re{U_{\alpha k} U^{*}_{\beta k} U^{*}_{\alpha j} U_{\beta j}}\sin^2{\frac{\Delta m^2_{kj}}{4E}L}+2\sum_{k>j}\Im{U_{\alpha k} U^{*}_{\beta k} U^{*}_{\alpha j} U_{\beta j}}\sin{\frac{\Delta m^2_{kj}}{2E}L}.
\end{equation}

Thus, sought-for $\Delta P$:

\begin{equation}
\Delta P= 4\sum_{k>j}\Im{U^*_{\alpha k} U_{\beta k} U_{\alpha k} U^{*}_{\beta j}}\sin{\frac{\Delta m^2_{kj}}{2E}L}. \label{eq5}
\end{equation}

The expression does not equal zero generally, which indicates that \textit{CP}-symmetry is broken in oscillations of three neutrino generations. It is already clear from \eqref{eq5} that the essential condition of the violation is the presence of complex elements in the mixing matrix, and that means the presence of the only Dirac phase in the chosen parameterization \eqref{eq2}. In order to see the conditions by which the \textit{CP}-symmetry recovers, let us write down the explicit expression \eqref{eq5} for $\nu_e$ t $\nu_{\mu}$ oscillations:

\begin{equation}
\Delta P_{\nu_{e}\rightarrow \nu_{\mu}}=4c_{12}c^2_{13}c_{23}s_{13}s_{23}s_{12}(\sin{\frac{\Delta m^2_{31}}{2E}L}-\sin{\frac{\Delta m^2_{21}}{2E}L}-\sin{\frac{\Delta m^2_{32}}{2E}L})\sin{\delta}. \label{eq6}
\end{equation}
Calculation of the value \eqref{eq6} is presented in Appendix B.

From this formula it is explicitly seen that $\Delta P=0$ and the \textit{CP}-violation vanishes by the following conditions: 1) the Dirac phase equals zero; 2) sine or cosine of any mixing angles equal zero; 3) any of two neutrino masses equal zero. It is experimentally known \cite{cit2}, that the mixing angles do not equal zero or $\pi/2$ and mass square differences do not equal zero, although they are very small. The second item implements only when there are two neutrino generations. In this case, the condition of the first item is satisfied too, because when the third generation is absent the \textit{CP}-violating phase in PMNS-matrix also absent. The most interesting item is the third, because it leads to the nontrivial consequences by taking into account all three generations: the recovery of \textit{CP}-symmetry in oscillations. Below there will be considered two explanations of this phenomenon.

\section{Oscillations in case of two degenerate mass states}

For the existence of the \textit{CP}-symmetry it is necessary that the amplitude of being studied process had at least two components with different \textit{CP}-invariant and \textit{CP}-noninvariant phases. In the present Section, it will be shown how \textit{CP}-symmetry recovers when two neutrino state masses are equal: in Section~3.1 the \textit{CP}-violating Dirac phase will be removed from PMNS-matrix; in Section 3.2 \textit{CP}-violating phase will be retained, but the violation will disappear because of the \textit{CP}-invariant phases equality.

\subsection{The removal of \textit{CP}-violating phase in the mixing matrix}
The mixing matrix represents a unitary matrix with $2n^2$ real parameters, among which $2n^2-n-2n(n-1)/2=n^2$ parameters are independent. Any unitary matrix can be parametrized with angles and phases. The number of the independent angles in this case coincides with the number of independent angles in orthogonal matrix from $O(n)$ group, i.e. $n^2-n-n(n-1)/2=n(n-1)/2$. Then the number of phases equals $n^2-n(n-1)/2=n(n+1)/2$. However by multiplying on the unobservable phases of left charged leptons and neutrino fields in the weak charged current, which is a part of the Standard model Lagrangian, we can remove in addition $2n-1$ phase in this matrix. Then there are $n(n+1)/2-(2n-1)=(n-2)(n-1)/2$ physical phases left. Hence it is clear that \textit{CP}-violating phase appears, when there are at least three generations. Let us consider that $n=3$. Then PMNS-matrix factorizes in standard parameterization in the following way:

\begin{equation}
U_{\mathrm{PMNS}}=R_{23}\Psi^{\dagger}R_{13}\Psi R_{12},
\end{equation}
where $R_{12}, R_{13}, R_{23}$ are real matrixes of rotation in the plane $1$---$2$ to the angle $\theta_{12}$, in the plane $1$---$3$ to the angle $\theta_{13}$, in the plane $2$---$3$ to the angle $\theta_{23}$ correspondingly, and $\Psi=\mathrm{diag}(1,1,e^{i\delta})$. In this expression unphysical phases are already removed by the corresponding rotations of the lepton fields. But for the further reasoning we need another factorization. Let us consider an arbitrary unitary $3\times3$-matrix.
Following \cite{cit3} it can be represented in the form of 

\begin{equation}
    U=D(\omega_1, \omega_2, \omega_3)W_{23}(\theta_{23}, \delta_{23})W_{13}(\theta_{13}, \delta_{13})W_{12}(\theta_{12}, \delta_{12}), \label{eq8}
\end{equation}
where $D(\omega_1, \omega_2, \omega_3)=\mathrm{diag}(e^{i\omega_1}, e^{i\omega_2}, e^{i\omega_3})$, $W_{ij}(\theta_{ij}, \delta_{ij})=\Phi(\delta_{ij})R(\theta_{ij})\Phi^{\dagger}(\delta_{ij})$, $R(\theta_{ij})$ is a matrix of rotation to angle $\theta_{ij}$ in the plane $i$---$j$. For example, for $i=1, j=2$ we have:

\begin{equation}
\Phi(\delta_{12})=\mathrm{diag}(e^{i\delta_{12}}, 1, 1),\
    R(\theta_{12})=
    \begin{pmatrix}
    \cos{\theta_{12}} & \sin{\theta_{12}} & 0 \\
    -\sin{\theta_{12}} & \cos{\theta_{12}} & 0 \\
    0 & 0 & 1 
    \end{pmatrix},\
    W_{12}=
    \begin{pmatrix}
    \cos{\theta_{12}} & e^{i\delta_{12}}\sin{\theta_{12}} & 0 \\
    -e^{-i\delta_{12}}\sin{\theta_{12}} & \cos{\theta_{12}} & 0 \\
    0 & 0 & 1 
    \end{pmatrix}.
\end{equation}

The order of multiplying matrices $W_{ij}$ in \eqref{eq8} can be chosen arbitrary. More general case of the unitary matrix $U_{PMNS}$ expansion by $W_{ij}$, which have three phases, will be considered in the Appendix C. Now let us assume for definiteness, that $|\nu_1\rangle$ and $|\nu_2\rangle$ are degenerate. Since we know, that there is only one physical phase, we will redistribute phases so that only one Dirac phase remains inside the product of matrices, and the other five phases are contained in matrices on the left and right, which will then be easily eliminated by rotating the lepton fields. For this purpose we will make the identical transformation:

\begin{equation}
\begin{gathered}
    U=D(\omega_1-\phi_1, \omega_2-\phi_2, \omega_3-\phi_3)\{ D(\phi_1, \phi_2, \phi_3)W_{23}(\theta_{23}, \delta_{23})D^{\dagger}(\phi_1, \phi_2, \phi_3)D(\phi_1, \phi_2, \phi_3)W_{13}(\theta_{13}, \delta_{13})\times \\ \times D^{\dagger}(\phi_1, \phi_2, \phi_3)D(\phi_1, \phi_2, \phi_3)W_{12}(\theta_{12}, \delta_{12})D^{\dagger}(\phi_1, \phi_2, \phi_3)\}D(\phi_1, \phi_2, \phi_3)=D(\omega_1-\phi_1, \omega_2-\phi_2, \omega_3-\phi_3)\times\\\times \{W_{23}(\theta_{23},\delta_{23}+\phi_1-\phi_2)W_{13}(\theta_{13},\delta_{13}+\phi_1-\phi_3)W_{12}(\theta_{12}, \delta_{12}+\phi_2-\phi_3)\}D(\phi_1, \phi_2, \phi_3),\label{eq10}
    \end{gathered}
\end{equation}
where we used, for example, for $W_{12}$: 
\begin{equation}
\begin{gathered}
D(\phi_1, \phi_2, \phi_3)W_{12}(\theta_{12}, \delta_{12})D^{\dagger}(\phi_1, \phi_2, \phi_3)=D(\phi_1, \label{eq11} \phi_2, \phi_3)\Phi(\delta_{12})R(\theta_{12})\Phi(-\delta_{12})D(-\phi_1, -\phi_2,-\phi_3)=\\=D(\delta_{12}+\phi_1, \phi_2, \phi_3) R(\theta_{12})D^{\dagger}(\phi_1+\delta_{12}, \phi_2, \phi_3)=W_{12}(\theta_{12}, \delta_{12}+\phi_1-\phi_2).
\end{gathered}
\end{equation}

Now there are nine phases in matrix $U$: $\omega_1-\phi_1$, $\omega_2-\phi_2$, $\omega_3-\phi_3$, $\delta_{12}+\phi_1-\phi_2$, $\delta_{13}+\phi_1-\phi_3$, $\delta_{23}+\phi_2-\phi_3$, $\phi_1$, $\phi_2$, $\phi_3$. In total, as it was shown at the beginning of the Section, there are six independent phases at unitary $3\times3$-matrix, then the remaining six phases are arbitrary. Obviously, that these are the phases $\phi_1, \phi_2, \phi_3$ introduced in \eqref{eq10}. Let us choose these phases in the following way: 

\begin{equation}
    \phi_2=\delta_{23}+\phi_1,\ \phi_3=\delta_{13}+\phi_1. \label{eq12}
\end{equation}

Then matrices $W_{23}, W_{13}$ are simply matrices of rotation $R_{23}, R_{13}$, and the matrix $W_{12}$ will keep the only phase $\delta_{12}+\delta_{23}-\delta_{13}$. There is one more undefined phase $\phi_1$ left. We assume it to be equal zero. Then taking into account \eqref{eq12} $D(\omega_1-\phi_1$, $\omega_2-\phi_2, \omega_3-\phi_3)=D(\omega_1, \omega_2-\delta_{23}$, $\omega_3-\delta_{13}), D(\phi_1, \phi_2, \phi_3)=D(1, \delta_{23}, \delta_{13})$. Now it is clear, that the introducing of three new phases in \eqref{eq10} leads to the reparameterization of the matrix $U$ through the factorization of five of six phases into the separate matrices $D(\overrightarrow{\phi})$. It remains to remember that the physical phase is the only one and the rest phases can be removed by the rotations of the charged leptons and neutrinos, so making this rotation we eliminate the phase matrices $D(\overrightarrow{\phi})$. Finally we have for the matrix $U$ in a weak charged current:

\begin{equation}
    U\equiv U_{\mathrm{PMNS}}=R_{23}(\theta_{23})R_{13}(\theta_{13})W_{12}(\theta_{12},\label{eq13} \delta_{\textit{CP}}),
\end{equation}
where we identified the rest phase in $W_{12}$ with the \textit{CP}-violating Dirac phase. As it was noticed, the matrices $W_{ij}$ multiplying order is an arbitrary, and so either of the matrices $W_{23}, W_{13}$ may ultimately remain on the right in expression \eqref{eq13} in place of the matrix $W_{12}$, depending on the order we will choose in expression \eqref{eq8}. Therefore the argumentation preceded above extends on the case of any other degenerate states. 

As the masses of the states $|\nu_1\rangle$ and $|\nu_2\rangle$ are equal, so there is an additional symmetry, namely, the symmetry of the unitary $U(2)$ rotations in the space $\nu_1$---$\nu_2$. Let us make such a transformation $|\nu_k\rangle \rightarrow V_{12}|\nu_k\rangle$, where $V_{12}$, where $V_{12}$ is a unitary matrix, which looks like the following:

\begin{equation}
V_{12}=
\begin{pmatrix}
v_{11} & v_{12} & 0\\
v_{21} & v_{22} & 0 \\
0 & 0 & 1
\end{pmatrix}.
\end{equation}

Such matrix has three phase and one angle. Now it is sufficient to substitute $U=R_{23}(\theta_{23})R_{13}(\theta_{13})\times\\ \times W_{12}(\theta_{12}, \delta_{\textit{CP}})\rightarrow R_{23}(\theta_{23})R_{13}(\theta_{13})W_{12}(\theta_{12}, \delta_{\textit{CP}}) V_{12}$ in the weak charged current. The unitary matrix $W_{12}(\theta_{12}, \delta_{\textit{CP}})$ is:
\begin{equation}
W_{12}(\theta_{12}, \delta_{\textit{CP}})=\begin{pmatrix}
\cos{\theta_{12}} & e^{i\delta_{\textit{CP}}}\sin{\theta_{12}} & 0\\
-e^{-i\delta_{\textit{CP}}}\sin{\theta_{12}} & \cos{\theta_{12}} & 0 \\
0 & 0 & 1
    \end{pmatrix}.
\end{equation}

Choosing the matrix $V_{12}$ so that $V_{12}=W^{\dagger}_{12}(\theta_{12}, \delta_{\textit{CP}})$, we will remove the Dirac phase, leaving the matrix $U_{\mathrm{PMNS}}$ completely real. Then \eqref{eq5} turns to zero, consequently, there is no \textit{CP}-violation. Remembering, that \textit{CP}-violating phase occurs in consequence of impossibility to eliminate all six phases by rotation of the charged lepton and neutrino fields , we conclude that extra $U(2)$-symmetry adds one missing phase to cancel the $\delta_{\textit{CP}}$.

\subsection{\textit{CP}-invariant and noninvariant phases in the amplitude}

Let us assume that we have not made a unitary rotation, thereby preserving the Dirac phase in the mixing matrix. We will show why nevertheless the \textit{CP}-violation vanishes. Let us consider the oscillations of electron neutrino with the definite momentum $p$ into a muonic one at the distance $L$ from the source and calculate explicitly the amplitude of oscillations:
\begin{equation}
|\nu_e(L,t)\rangle=e^{-iE_1t+ip_1L}U_{e1}|\nu_1\rangle+e^{-iE_2t+ip_2L}U_{e2}|\nu_2\rangle+e^{-iE_3t+ip_3L}U_{e3}|\nu_3\rangle,
\end{equation}

\begin{equation}
|\nu_{\mu}\rangle=U_{\mu1}|\nu_1\rangle+U_{\mu1}|\nu_2\rangle+U_{\mu3}|\nu_3\rangle.
\end{equation}

\begin{equation}
\begin{gathered}
A_{\nu_{e}\rightarrow{\nu_{\mu}}}=\langle\nu_{\mu}|\nu_e(L,t)\rangle=e^{ipL}(e^{-iE_1t}U^{\dagger}_{\mu1}\label{eq18} U_{e1}+e^{-iE_2t}U^{\dagger}_{\mu2}U_{e2}+e^{-iE_3t}U^{\dagger}_{\mu3}U_{e3})\approx\\\approx e^{ipL}(e^{-i(p+\frac{m^2_1}{2p})L}U^{\dagger}_{\mu1}U_{e1}+e^{-i(p+\frac{m^2_2}{2p})L}U^{\dagger}_{\mu2}U_{e2}+e^{-i(p+\frac{m^2_3}{2p})L}U^{\dagger}_{\mu3}U_{e3})\approx\\ \approx e^{-i\frac{m^2_1}{2E}L}U^{\dagger}_{\mu1}U_{e1}+e^{-i\frac{m^2_2}{2E}L}U^{\dagger}_{\mu2}U_{e2}+e^{-i\frac{m^2_3}{2E}L}U^{\dagger}_{\mu3}U_{e3},
\end{gathered}
\end{equation}
where we used the fact that neutrino is ultrarelativistic. i.e. $E\approx |p|$ and $t\approx L$.

As mentioned before when it is about the \textit{CP}-violation, there always must be invariant and noninvariant phases with respect to \textit{CP}-transformation. In the expression for the amplitude of oscillations \eqref{eq18} the phases in exponents attached to the elements of the mixing matrix are \textit{CP}-invariant, and the phases in the elements themselves are \textit{CP}-noninvariant. Therefore by \textit{CP}-transformation in the amplitude there is only substitution $U^{\dagger}_{\mu i} U_{ei} \rightarrow U^{\dagger}_{ei} U_{\mu i}$. Let us use the unitarity of PMNS-matrix:
\begin{equation}
U^{\dagger}_{\mu1}U_{e1}+U^{\dagger}_{\mu2}U_{e2}+U^{\dagger}_{\mu3}U_{e3}=0. \label{eq19}
\end{equation}

We assume that the states $|\nu_1\rangle$ and $|\nu_2\rangle$ are degenerate, i.e. $m_1=m_2$. Then the expressions \eqref{eq18} and \eqref{eq19} lead to:
\begin{equation}
\begin{gathered}
A_{\nu_{e}\rightarrow{\nu_{\mu}}}=(e^{-i\frac{m^2_1}{2E}L}-e^{-i\frac{m^2_3}{2E}L})U^{\dagger}_{\mu1}U_{e1}+\label{eq20} (e^{-i\frac{m^2_2}{2E}L}-e^{-i\frac{m^2_3}{2E}L})U^{\dagger}_{\mu2}U_{e2}=2\sin{\frac{\Delta m^2_{31}}{4E}L}\bigg(\sin{\frac{m^2_1+m^2_3}{4E}L}+\\+i\cos{\frac{m^2_1+m^2_3}{4E}L}\bigg)U^{\dagger}_{\mu1}U_{e1}+2\sin{\frac{\Delta m^2_{32}}{4E}L}\bigg(\sin{\frac{m^2_3+m^2_2}{4E}L}+i\cos{\frac{m^2_3+m^2_2}{4E}L}\bigg)U^{\dagger}_{\mu2}U_{e2}=\\=2U^{\dagger}_{\mu1}U_{e1}e^{-i(\frac{m^2_3+m^2_1}{4E}L-\frac{\pi}{2})}\sin{\frac{\Delta m^2_{31}}{4E}L}+2U^{\dagger}_{\mu2}U_{e2}e^{-i(\frac{m^2_3+m^2_2}{4E}L-\frac{\pi}{2})}\sin{\frac{\Delta m^2_{32}}{4E}L}.
\end{gathered}
\end{equation}

There are two components in expression for the amplitude \eqref{eq20} with the different \textit{CP}-noninvariant, and identical \textit{CP}-invariant phases. Taking into account \eqref{eq19} the probability of the oscillations equals:
\begin{equation} 
P_{\nu_{e}\rightarrow{\nu_{\mu}}}=|A_{\nu_{e}\rightarrow{\nu_{\mu}}}|^2=4|U_{\mu 3}|^2|U_{e3}| \label{eq21}^2\sin^2{\frac{\Delta m^2_{32}}{4E}L}.
\end{equation}

Under the \textit{CP}-transformation the amplitude \eqref{eq20} turns into 
\begin{equation}
A_{\overline{\nu_{e}}\rightarrow \overline{{\nu_{\mu}}}}=2e^{-i(\frac{m^2_3+m^2_2}{4E}L-\frac{\pi}{2})} \label{eq22} \sin{\frac{\Delta m^2_{32}}{4E}L}\big(U_{\mu1}U^{\dagger}_{e1}+U_{\mu2}U^{\dagger}_{e2}\big),
\end{equation}
where the equal \textit{CP}-preserving phase is extracted as a common factor. It means that $P_{\overline{\nu_{e}}\rightarrow \overline{{\nu_{\mu}}}}=|A_{\overline{\nu_{e}}\rightarrow \overline{{\nu_{\mu}}}}|^2=P_{\nu_{e}\rightarrow{\nu_{\mu}}}$, i.e. \textit{CP}-symmetry has recovered despite the presence of the \textit{CP}-violating phase and its difference for both terms in the amplitudes \eqref{eq20}, \eqref{eq22}. It is interesting to notice that formula \eqref{eq21} is absolutely analogous to the expression of the probability of the oscillations in case of two neutrino generations. The degeneracy of the generations $|\nu_1\rangle$ and $|\nu_2\rangle$ is explicitly reflected here, as the substitution $\Delta m^2_{32}\rightarrow\Delta m^2_{31}$ does not change the probability of oscillations, and the factor $|U_{\mu 3}|^2|U_{e3}|^2$ relates to the generation $|\nu_3\rangle$. Thus, in terms of oscillations there are two states namely nondegenerate $|\nu^{'}_2\rangle=|\nu_3\rangle$ and twofold degenerate by mass $|\nu^{'}_1\rangle$, which takes on values $|\nu_1\rangle$, $|\nu_2\rangle$. And as it was mentioned the \textit{CP}-violation exists only when there are at least three different neutrino generations, therefore the violation is absent in this case.

\section{Conclusion}

In the present article there are \textit{CP}-symmetry recovery mechanisms in case of three neutrino generations with two degenerate mass states demonstrated. The first method allows to eliminate the \textit{CP}-violating phase in the amplitude of oscillations with the $U(2)$-transformation in the space of the degenerate states. In the second method the degeneracy of two mass states let us to factorize the amplitude by separating \textit{CP}-invariant and \textit{CP}-noninvariant factor, which leads to the equality of the squared absolute values of the amplitudes of the direct and \textit{CP}-transformed processes, i.e. the equality of the probabilities of this processes.

Since we know \cite{cit2}, that $\Delta m^2_{12}=(7.5\pm 0.2)\times 10^{-5}\ $eV$^2$, $\Delta m^2_{23}\approx \Delta m^2_{13}=(2.5\pm 0.2)\times 10^{-3} \ $eV$^2$, so the approximation of the equivalence of two neutrino mass states is quite legitimate. From \eqref{eq6} it leads that $\Delta P\sim \frac{|\Delta m^2_{12}|}{2E}L$, and so the \textit{CP}-violation effect is small on the distances $L\lesssim \frac{E}{\Delta m^2_{12}}$.

It is worth noting that \textit{CP}-violation is a necessary condition for the baryon asymmetry generation. The effect of the \textit{CP}-violation in the quark sector is defined by the similar expression to the case in the neutrino sector \cite{cit4}: $d_{\textit{CP}}=\sin{\theta_{12}}\sin{\theta_{23}}\sin{\theta_{13}}\sin{\delta_{\textit{CP}}}(m^2_{t}-m^2_{c})(m^2_{t}-m^2_{u})(m^2_{c}-m^2_{u})(m^2_{b}-m^2_{s})(m^2_{b}-m^2_{d})(m^2_{s}-m^2_{d})$, where the degeneracy of two quarks of the up sort and/or the down sort leads to the \textit{CP}-symmetry recovery. However in contrast to the effect of the \textit{CP}-violation in the neutrino oscillations, the effect is very small in the quark case and therefore it is not appropriate for the explanation of the observable asymmetry \cite{cit5}: actually, for the temperature of the electroweak phase transition $T$ which is in the order of hundreds GeV we obtain: $d_{\textit{CP}}/T^{12}\simeq 10^{-20}$. 

When deriving the expression for the amplitude of oscillations, we considered, that neutrinos are ultrarelativistic and have a definite momentum (see Appendix A). In general we should work with the model of the wave packet when we consider the neutrino oscillations in vacuum. In this case the neutrino momentum is distributed about some average value with the definite dispersion. The probability of oscillations takes then more complicated form as in formulas \eqref{eq10}, \eqref{eq11} in the work \cite{cit6}. It is essential, that each term in the expression for the probability consists of the \textit{CP}-noninvariant factor in the form of the product of the mixing matrix elements and \textit{CP}-invariant factor, depending on the neutrino masses square difference only and the values describing the wave packet. Therefore by the \textit{CP}-transformation only \textit{CP}-noninvariant factors are replaced by complex-conjugated ones, and for the magnitude of the \textit{CP}-violation effect we obtain the expression similar to \eqref{eq5} and differing from it only in \textit{CP}-invariant factors. So the reasoning, presented in the Section 3, are applied in this case consequently when two neutrino states are degenerate the \textit{CP}-symmetry recovers. Thus the consideration of the wave packet case does not make any qualitative changes in the question about \textit{CP}-violation in the neutrino oscillations, but introduces additional parameters of the wave packet into the \textit{CP}-invariant part of the oscillation probability. 

Finally, we will dwell upon how the Dirac versus Majorana neutrino-mass dichotomy affects \textit{CP}-symmetry restoration. In Section 2, it was stated that the mixing matrix $U_{PMNS}$ involves only one physical \textit{CP}-violating phase. This corresponds to the assumption that the neutrino is a Dirac particle. Let us now assume that the neutrino is a Majorana particle. This means that this particle coincides with its antiparticle, in which case the mixing matrix contains two phases in addition to the Dirac phase. As before, we assume that some two states are degenerate. The method proposed in Subsection 3.1 for removing \textit{CP}-violation becomes inapplicable, since this method is able to eliminate only one phase in the Majorana neutrino mixing matrix. However, the argument in Subsection 3.2 remains valid, since it relies on the equality of the \textit{CP}-invariant phases in the amplitude and therefore does not change in the presence of additional \textit{CP}-noninvariant phases in the mixing matrix. Thus, the degeneracy of states leads to \textit{CP}-symmetry restoration in the case of Majorana neutrinos inclusive.

The author thanks his scientific advisor, M.I. Vysotsky for the formulation of the problem and fruitful discussions, and S.I. Godunov for the discussions and constructive criticism during the work on the article. This article is backed by the Russian Science Foundation grant No 19-12-00123.

\section*{\textit{Appendix A}}
The derivation of the expression for the probability of neutrino oscillations in case of three neutrino generations is based on the work \cite{cit2}. Let at the initial moment of time there is a neutrino with flavor $\alpha$ and momentum $p$. The propagation of the neutrino mass state in time and space in the context of the considering problem may be described by a plane wave: $|\nu_k(t,x)\rangle=e^{-iE_{k}t+ip_{k}x}|\nu_{k}\rangle$, where $k=1,2,3$, $E_k=\sqrt{p^2_k+m^2_k}$. Knowing the mixing matrix we can write for the evolution of the flavour state\footnote{In the work the summation is written explicitly}: 
\begin{equation}
|\nu_{\alpha}(t, x)\rangle = \sum_k U_{\alpha k}|\nu_k(t, x)\rangle=\sum_k U_{\alpha k}e^{-iE_{k}t+ip_{k}x}|\nu_k\rangle,\ \alpha=e, \mu, \tau. \tag{A.1}
\end{equation}

Since we consider the neutrino to have a definite momentum, $p_i=p$ for $i=1,2,3$. There is an opportunity to choose the other initial condition, namely the birth of neutrino with the definite energy $E$. However in the ultrarelativistic limit the both choices lead to the same result \cite{cit7}. We get the amplitude of the oscillations:

\begin{equation}
A_{\nu_{\alpha}\rightarrow{\nu_{\beta}}}=\langle\nu_{\beta}|\nu_{\alpha}(t,x) \rangle=\sum_k U_{\alpha k} U^{*}_{\beta k}e^{-iE_{k}t+ipx}. \tag{A.2} 
\end{equation}

Then the probability of oscillations is equal:

\begin{equation}
P_{\nu_{\alpha}\rightarrow{\nu_{\beta}}}=|A_{\nu_{\alpha}\rightarrow{\nu_{\beta}}}|^2=\sum_{k,j} U_{\alpha k} U^{*}_{\beta k} U^{*}_{\alpha j} U_{\beta j} e^{-i(E_k-E_j)t}. \tag{A.3} \label{eqa3}
\end{equation}

Considering the neutrino ultrarelativistic, i.e. 
$m\ll p$, we obtain

\begin{equation}
(E_k-E_j)t=\Big(\sqrt{p^2+m^2_k}-\sqrt{p^2+m^2_j}\Big)t\approx \Big(p(1+\frac{m^2_k}{2p^2})-p(1+\frac{m^2_j}{2p^2})\Big)t=\frac{\Delta m^2_{kj}}{2p}t\approx \frac{\Delta m^2_{kj}}{2E}L, \tag{A.4} \label{eqa4}
\end{equation}
where $\Delta m^2_{kj}\equiv m^2_k-m^2_j$ and it is used, that in the ultrarelativistic case $E\simeq p$, $t\simeq L$.

Let us separate explicitly the constant term from from the oscillating term in the expression \eqref{eqa3}:

\begin{equation}
\sum_{k,j} U_{\alpha k} U^{*}_{\beta k} U^{*}_{\alpha j} U_{\beta j} e^{-i(E_k-E_j)t}=\sum_k |U_{\alpha k}|^2|U_{\beta k}|^2+2\sum_{k>j}\Re{U_{\alpha k} U^{*}_{\beta k} U^{*}_{\alpha j} U_{\beta j} e^{-i\frac{\Delta m^2_{kj}}{2E}}L}, \tag{A.5} \label{eqa5}
\end{equation}
where we used \eqref{eqa4}, and in the first term for $k=j$ we have $e^{-i(E_k-E_j)t}=1$. By unitarity of the PMNS-matrix $\sum_k U_{\alpha k} U^{*}_{\beta k}=\sum_j U^{*}_{\alpha j} U_{\beta j}=\delta_{\alpha \beta}$, we obtain

\begin{equation}
\begin{gathered}
\delta_{\alpha \beta}\cdot \delta_{\alpha \beta}=(\sum_k U_{\alpha k} U^{*}_{\beta k})(\sum_j U^{*}_{\alpha j} U_{\beta j})=\sum_{k,j} U_{\alpha k} U^{*}_{\beta k} U^{*}_{\alpha j} U_{\beta j}=\sum_{k=j} U_{\alpha k} U^{*}_{\alpha j} U^{*}_{\beta k} U_{\beta j}+\sum_{k\neq j} U_{\alpha k} U^{*}_{\beta k} U^{*}_{\alpha j} U_{\beta j}=\\=\sum_k |U_{\alpha k}|^2|U_{\beta k}|^2+\sum_{k>j} U_{\alpha k} U^{*}_{\beta k} U^{*}_{\alpha j} U_{\beta j}+\sum_{k>j}U_{\alpha j} U^{*}_{\beta j} U^{*}_{\alpha k} U_{\beta k}=\sum_k |U_{\alpha k}|^2|U_{\beta k}|^2+\sum_{k>j} U_{\alpha k} U^{*}_{\beta k} U^{*}_{\alpha j} U_{\beta j}+\\+\sum_{k>j}(U_{\alpha k} U^{*}_{\beta k} U^{*}_{\alpha j} U_{\beta j})^{*}=\sum_k |U_{\alpha k}|^2|U_{\beta k}|^2+2\sum_{k>j}\Re{U_{\alpha k} U^{*}_{\beta k} U^{*}_{\alpha j} U_{\beta j}}=\delta_{\alpha \beta}, \tag{A.6} 
\end{gathered}
\end{equation}

Dividing the exponent \eqref{eqa5} into the real and imaginary parts, eventually we get:

\begin{equation}
\begin{gathered}
P_{\nu_{\alpha}\rightarrow{\nu_{\beta}}}=\delta_{\alpha \beta}-2\sum_{k>j}\Re{U_{\alpha k} U^{*}_{\beta k} U^{*}_{\alpha j} U_{\beta j}}(1-\cos{\frac{\Delta m^2_{kj}}{2E}}L)-2\sum_{k>j}\Im{U_{\alpha k} U^{*}_{\beta k} U^{*}_{\alpha j} U_{\beta j}}\sin{\frac{\Delta m^2_{kj}}{2E}L}=\\=\delta_{\alpha \beta}-4\sum_{k>j}\Re{U_{\alpha k} U^{*}_{\beta k} U^{*}_{\alpha j} U_{\beta j}}\sin^2{\frac{\Delta m^2_{kj}}{4E}L}-2\sum_{k>j}\Im{U_{\alpha k} U^{*}_{\beta k} U^{*}_{\alpha j} U_{\beta j}}\sin{\frac{\Delta m^2_{kj}}{2E}L}. \tag{A.7} 
\end{gathered}
\end{equation}

\section*{\textit{Appendix B}}

Let us obtain the explicit form of the expression \eqref{eq5} for the oscillations $\nu_e\rightarrow \nu_{\mu}$:

\begin{equation}
\begin{gathered}
    \Delta P_{\nu_{e}\rightarrow \nu_{\mu}}=4\Im{U^{*}_{\mu2} U_{e2} U_{\mu1} U^{*}_{e1}}\sin{\frac{\Delta m^2_{21}}{2E}L}+4\Im{U^{*}_{\mu 3} U_{e3} U_{\mu1} U^{*}_{e1}}\sin{\frac{\Delta m^2_{31}}{2E}L}+\\+4\Im{U^{*}_{\mu 3} U_{e3} U_{\mu 2} U^{*}_{e2}}\sin{\frac{\Delta m^2_{32}}{2E}L}. \tag{B.1} 
\end{gathered}
\end{equation}

We write down the elements attached to the each $\sin{\frac{\Delta m^2_{kj}}{2E}L}$:

\begin{equation}
\begin{gathered}
\Delta m^2_{21}: U^{*}_{\mu2} U_{e2} U_{\mu1} U^{*}_{e1}=(c_{12}c_{23}-s_{12}s_{23}s_{13}e^{-i\delta})s_{12}c_{13}(-s_{12}c_{23}-c_{12}s_{23}s_{13}e^{i\delta})c_{12}c_{13}=\\=(-c_{12}c^2_{23}s_{12}-c^2_{12}c_{23}s_{23}s_{13}e^{i\delta}+s^2_{12}c_{23}s_{23}s_{13}e^{-i\delta}+s_{12}c_{12}s^2_{23}s^2_{13})=\\=c_{12}c^2_{13}s_{12}(c_{23}s_{23}s_{13}(s^2_{12}e^{-i\delta}-c^2_{12}e^{i\delta})+s_{12}c_{12}((s_{23}s_{13})^2-c^2_{23}))\Longrightarrow\\ \Longrightarrow \Im{U^{*}_{\mu2} U_{e2} U_{\mu1} U^{*}_{e1}}=-c^2_{13}c_{12}c_{23}s_{13}s_{12}s_{23}\sin{\delta}; \tag{B.2} 
\end{gathered}
\end{equation}

\begin{equation}
\begin{gathered}
\Delta m^2_{31}: U^{*}_{\mu 3} U_{e3} U_{\mu1} U^{*}_{e1}=s_{23}c_{13}s_{13}e^{-i\delta}(-s_{12}c_{23}-c_{12}s_{23}s_{13}e^{i\delta})c_{12}c_{13}= \\ \nonumber =c^2_{13}c_{12}s_{23}s_{13}(-s_{12}c_{23}e^{-i\delta}-c_{12}s_{23}s_{13})\Longrightarrow\Im{U^{*}_{\mu 3} U_{e3} U_{\mu1} U^{*}_{e1}}=c^2_{13}c_{12}c_{23}s_{13}s_{12}s_{23}\sin{\delta}; \tag{B.3} 
\end{gathered}
\end{equation}

\begin{equation}
\begin{gathered}
\Delta m^2_{32}: U^{*}_{\mu 3} U_{e3} U_{\mu 2} U^{*}_{e2}=s_{23}c_{13}s_{13}e^{-i\delta}(c_{12}c_{23}-s_{12}s_{23}s_{13}e^{i\delta})s_{12}c_{13}=\\ \nonumber =s_{23}c^2_{13}s_{13}s_{12}(c_{12}c_{23}e^{-i\delta}-s_{12}s_{23}s_{13})\Longrightarrow\Im{U^{*}_{\mu 3} U_{e3} U_{\mu 2} U^{*}_{e2}}=-c^2_{13}c_{12}c_{23}s_{13}s_{12}s_{23}\sin{\delta}. \tag{B.4} 
\end{gathered}
\end{equation}

Finally we obtain for the conversion $\nu_e$ to $\nu_{\mu}$:

\begin{equation}
\Delta P_{\nu_{e}\rightarrow \nu_{\mu}}=4c_{12}c^2_{13}c_{23}s_{13}s_{23}s_{12}(\sin{\frac{\Delta m^2_{31}}{2E}L}-\sin{\frac{\Delta m^2_{21}}{2E}L}-\sin{\frac{\Delta m^2_{32}}{2E}L})\sin{\delta}. \tag{B.5} 
\end{equation}

\section*{\textit{Appendix C}}
We will show that an arbitrary unitary $3\times 3$-matrix of mixing can be made real when two neutrino mass states are degenerate. For this, we will consider its expansion to the matrices $W_{ij}$: $U=W_{12}W_{23}W_{13}$. The order of the product may be chosen arbitrary as in the Section 3.1. The matrices $W_{ij}$ may be parameterized with one real angle and three complex phases. Let us take for example $W_{12}(\theta_{12}, \phi_1, \psi, \delta)$:

\begin{equation}
W_{12}=e^{i\phi_1/2}
\begin{pmatrix}
e^{i\psi}\cos{\theta_{12}} & e^{i\delta}\sin{\theta_{12}} & 0\\
-e^{-i\delta}\sin{\theta_{12}} & e^{-i\psi}\cos{\theta_{12}} & 0 \\
0 & 0 & 1 \tag{C.1} 
\end{pmatrix}.
\end{equation}

Defining $\psi=\psi_1+\delta_1$, $\delta=\psi_1-\delta_1$, let us write the matrix $W_{12}(\theta_{12}, \phi_1, \psi, \delta)=W_{12}(\theta_{12}, \phi_1, \psi_1, \delta_1)$ in the following way:

\begin{equation}
W_{12}=e^{i\phi_1/2}
\begin{pmatrix}
e^{i\psi_1} & 0 & 0\\
0 & e^{-i\psi_1}& 0 \\
0 & 0 & 1
\end{pmatrix}
\begin{pmatrix}
\cos{\theta_{12}} & \sin{\theta_{12}} & 0\\
-\sin{\theta_{12}} & \cos{\theta_{12}} & 0 \\
0 & 0 & 1
\end{pmatrix}
\begin{pmatrix}
e^{i\delta_1} & 0 & 0\\
0 & e^{-i\delta_1} & 0 \\
0 & 0 & 1 \tag{C.2} \label{eqc2}
\end{pmatrix}.
\end{equation}

Thus, we have separated two phase matrices and one rotation matrix around the third axis. Parameterization of matrices $ W_{23} (\theta_{23}, \phi_2, \psi_2, \delta_2) $, $ W_{13} (\theta_{13}, \phi_3, \psi_3, \delta_3) $ is carried out in a similar way. As it was mentioned, the unitary matrix $3\times 3$ has three independent phases and three independent angles. In our expansion of the matrix $U$ there are nine phases and three angles. Hence we conclude that three phases are arbitrary. We substitute \eqref{eqc2} into the initial expansion of the matrix $U$, denoting the rotation matrices $ \theta_{12}, \theta_{23}, \theta_{13} $ as $ R_{12}, R_{23}, R_{ 13} $ respectively:

\begin{equation}
\begin{gathered}
U=e^{i(\phi_1/2+\phi_2/2+\phi_3/2)}
\begin{pmatrix}
e^{i\psi_1} & 0 & 0\\
0 & e^{-i\psi_1} & 0 \\
0 & 0 & 1 
\end{pmatrix}
R_{12}
\begin{pmatrix}
e^{i\delta_1} & 0 & 0\\
0 & e^{-i\delta_1} & 0 \\
0 & 0 & 1
\end{pmatrix}
\begin{pmatrix}
1 & 0 & 0\\
0 & e^{i\psi_2} & 0 \\
0 & 0 & e^{-i\psi_2}
\end{pmatrix}
R_{23}
\begin{pmatrix}
1 & 0 & 0\\
0 & e^{i\delta_2} & 0 \\
0 & 0 & e^{-i\delta_2}
\end{pmatrix}\times\\ \\ \times
\begin{pmatrix}
e^{i\psi_3} & 0 & 0\\
0 & 1 & 0 \\
0 & 0 & e^{-i\psi_3}
\end{pmatrix}
R_{13}
\begin{pmatrix}
e^{i\delta_3} & 0 & 0\\
0 & 1 & 0 \\
0 & 0 & e^{-i\delta_3}
\end{pmatrix}. \tag{C.3} 
\end{gathered}
\end{equation}

Let us assume $\psi_2=\delta_1=\delta_2=0$ and define $\Delta_1=\psi_3+\delta_3$, $\Delta_2=\psi_3-\delta_3$. Then for the matrix $U$ we have:

\begin{equation}
\begin{gathered}
U=e^{i(\phi_1/2+\phi_2/2)}
\begin{pmatrix}
e^{i\psi_1} & 0 & 0\\
0 & e^{-i\psi_1} & 0 \\
0 & 0 & 1 
\end{pmatrix}
R_{12}
R_{23}
e^{i\phi_3/2}
\begin{pmatrix}
e^{i\Delta_1}\cos{\theta_{13}} & 0 & e^{i\Delta_2}\sin{\theta_{13}}\\
0 & 1 & 0 \\
-e^{-i\Delta_2}\sin{\theta_{13}} & 0 & e^{-i\Delta_1}\cos{\theta_{13}}
\end{pmatrix}=\\=e^{i(\phi_1/2+\phi_2/2)}
\begin{pmatrix}
e^{i\psi_1} & 0 & 0\\
0 & e^{-i\psi_1} & 0 \\
0 & 0 & 1
\end{pmatrix}
R_{12}
R_{23}
W_{13}(\theta_{13}, \phi_3, \Delta_1, \Delta_2). \tag{C.4} 
\end{gathered}
\end{equation}

\end{document}